\documentclass[10pt,b5paper,twoside]{padeu}  
\usepackage{epsfig,natbib_padeu}             

\begin{document}

\title{Metallicity dependence of some parameters of Cepheids}
\titlerunning{Metallicity dependence of Cepheid properties}
\author{P. Klagyivik\inst{1}, 
	L. Szabados\inst{2}}
\authorrunning{P. Klagyivik and L. Szabados}
\institute{\inst{1} E\"otv\"os University, Department~of Astronomy, H-1518 Budapest, Pf.~32, Hungary \\
	   \inst{2} Konkoly Observatory of HAS, H-1525 Budapest, Pf.~67, Hungary
}
\email{\inst{1}P.Klagyivik@astro.elte.hu, 
       \inst{2}szabados@konkoly.hu}


\abstract{
Dependence of phenomenological properties of Cepheids on the heavy element
abundance is studied. It is found that the amplitude of the pulsation depends
on the metallicity of the stellar atmosphere.
}

\maketitle


\section{Introduction}

Metallicity has a very important role in Cepheids. Nowadays its influence on
the PLC relation is a frequently discussed topic of research papers. But we
only know little or nothing about other effects of metallicity.
\citet{Kovacs-etal:A&A95} and \citet{Jurcsik-etal:A&A96} worked out a method
to determine the value of $[Fe/H]$ from the shape of the light curves of
RR Lyrae stars. \citet{Zsoldos:ASPC95} tested this method for Cepheids too,
and found a similar relation.

An extensive project has been initiated for studying the effect of heavy
element abundance of Cepheids on various pulsational properties of these
radially oscillating stars. Here we present the first results on the
metallicity dependence of the photometric and radial velocity amplitudes.

\section{Data collection}

We collected the periods, the amplitudes of the light curves in
different colours, radial velocity curves and metallicity of galactic Cepheids.
Thanks to the recently aroused great interest there are about 150 Cepheids in
our galaxy with known $[Fe/H]$ value. The periods were taken from the DDO Database
\citep{Fernie-etal:IBVS95}, the amplitudes from three catalogues
(\citet{Fernie-etal:IBVS95}, \citet{Szabados:97},
\citet{Berd-etal:a&as00}) and the metallicities from the following publications:
\citet{Eggen:AJ85}, \citet{Giri:JApA86}, \citet{Fry-etal.:AJ97}
, \citet{Groene-etal:A&A04}, Andrievsky et al. (2002a, 2002b, 2002c)
, Andrievsky et al. (2004, 2005), \citet{Luck-etal:A&A03}, \citet{Kovty-etal:AJ05}.

In some cases the amplitudes of the light curves had quite different values as
determined by different authors. Thus we omitted the stars, if the differences
were larger than $0.1$ mag. For stars having several published metallicity values,
we averaged these data.

\section{Metallicity dependence}

\subsection{Photometric amplitude -- [Fe/H]}

First we studied the period--amplitude diagram. \citet{Kraft:ApJ60} explored
first this relation and he determined an upper envelope to the points on the
$\log P$ -- $\Delta B$ and $\log P$ -- $\Delta V$ diagrams.
\citet{Eichen-etal:A&A77} derived a method for constructing envelopes to point
diagrams and calculated an upper envelope for $\Delta B$. Their sample
contained 255 galactic Cepheids.

\begin{figure}[htb!]
   \centering
   \includegraphics[width=0.70\linewidth,angle=-90]{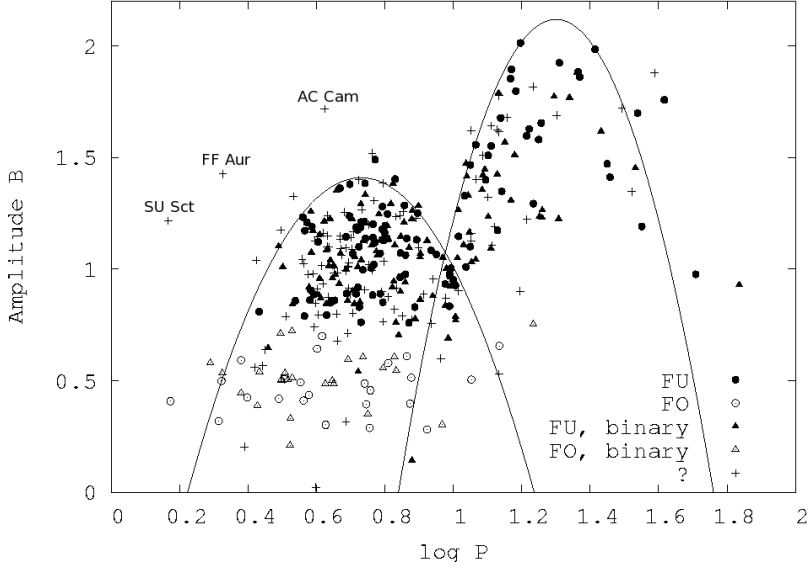}
   \caption{Upper envelope for $\Delta B$. The two curves are the fitted parabolas.
            FU: fundamental mode, FO: first overtone, ?: we have no information
	    about pulsation mode. The three stars marked with their names may
	    not be classical Cepheids.}
   \label{fig:burkolo}
   \end{figure}

Because more comprehensive catalogues have been published since then, we could
repeat this study by fitting two parabolae (Fig. \ref{fig:burkolo}).
These curves fitted better the dip near $\log P = 1$ than polynomials.
The equations of the parabolae are given in Eq.(1) and (2).
There are three stars (FF Aur, AC Cam, SU Sct) that lie far above the
envelope. It is possible, that they are not classical Cepheids.

\begin{eqnarray}
	 A_B = 1.41 - 5.5 \times (\log P - 0.73)^2 && \log P < 1\\
	 A_B = 2.15 - 10.0 \times (\log P - 1.3)^2 && \log P > 1
\end{eqnarray}

We investigated the metallicity dependence of the deviation of the amplitudes
from the estimated maximum amplitude ($\Delta B_{max} - \Delta B_*$) for the
given period. If a star has a companion, its light variation decreases because
of the smaller variation of the intensity ratios. First overtone pulsators also
have lower amplitudes. If only the solitary Cepheids (at least those without
any known companion), pulsating in the fundamental mode are plotted, there is a
definite trend (Figure 2). In the case of higher metallicity the amplitude
decreases compared to the upper envelope (i.e. the maximum possible value).
The equation of the straight line fit is given in Eq.(3) where DEV is the
deviation from the upper envelope.

   \begin{figure}[htb!]
   \centering
   \includegraphics[width=0.40\linewidth,angle=-90]{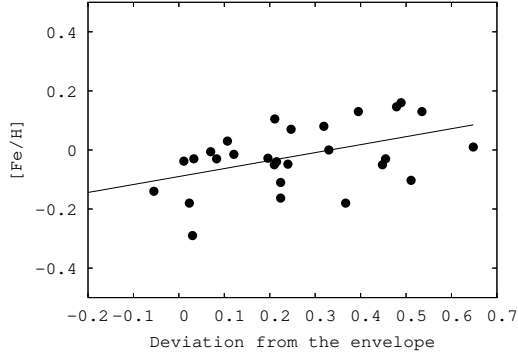}
   \caption{Deviation from the envelope for solitary, fundamental mode Cepheids.}
   \label{fig:elteres}
   \end{figure}

\begin{equation}
	[Fe/H] = -0.09 (\pm 0.03) + 0.27 (\pm 0.10) \times DEV
\end{equation}

\subsection{Amplitude ratio -- [Fe/H]}

\begin{figure}[htb!]
   \centering
   \includegraphics[width=0.40\linewidth,angle=-90]{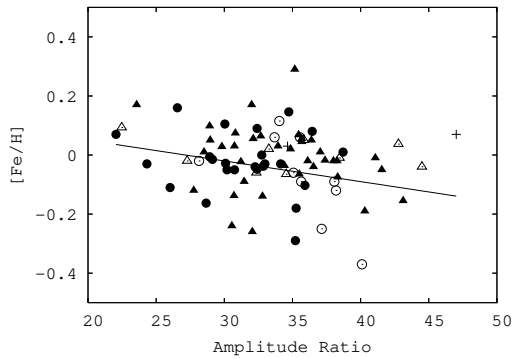}
   \caption{Metallicity dependence of the amplitude ratio. Symbols are the same
            as in Fig. \ref{fig:burkolo}.}
   \label{fig:fe-ar}
   \end{figure}

We defined the amplitude ratio as the ratio of the radial velocity and the
$B$ band photometric amplitude ($AR = \Delta V_{rad} / \Delta B$).
The relation between metallicity and AR and the linear fit is presented
in Figure \ref{fig:fe-ar} and Eq.(4). The fit applies only to the solitary
and fundamental mode Cepheids, but all the other groups show similar relation.

\begin{equation}
	[Fe/H] = 0.19 (\pm 0.10) - 0.007 (\pm 0.003) \times AR 
\end{equation}

\subsection{Slope parameter -- [Fe/H]}

If we plot the amplitude of the light variations in different colors compared
to the amplitude measured in $B$ band vs. $(1 / \lambda)$, the distribution
of the points will be roughly linear \citep{Fernie:PASP79}. \citet{Szabados:97}
defined the slope parameter as the slope of the straight line fitted to
these points.

   \begin{figure}[htb!]
   \centering
   \includegraphics[width=0.40\linewidth,angle=-90]{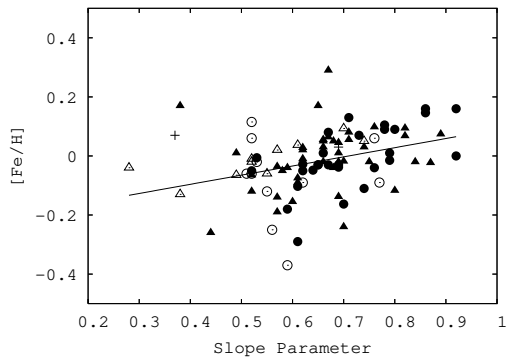}
   \caption{Metallicity dependence of the slope parameter. Symbols are the same
            as before.}
   \label{fig:fe-sp}
   \end{figure}

Fig. \ref{fig:fe-sp} shows the relation between metallicity and the slope parameter.
It is visible that the metallicity affects the slope of the line. The higher
the [Fe/H] the larger the slope parameter, i.e. the amplitude decreases quicker with
the increase of the wavelength. Probably this is due to the change (of the strength)
of the mechanisms that govern the pulsation. The result of the linear fit is given
in Eq.(5).

\begin{equation}
	[Fe/H] = -0.22 (\pm 0.06) + 0.31 (\pm 0.10) \times SP 
\end{equation}
where SP is the slope parameter. The fit applies only to the solitary and
fundamental mode Cepheids again, but all the other groups show similar
behaviour and the differences are not significant.

\section{Summary}

We analysed the metallicity dependence of three parameters related to the light
variation and radial velocity amplitudes of galactic Cepheids. We showed that
all these parameters (deviation from the upper envelope on the $\log P$--$\Delta B$
diagram, amplitude ratio and slope parameter) show a definite relation with $[Fe/H]$.
These are not exactly defined relations, rather tendencies (note that the errors
of $[Fe/H]$ are generally $0.05 - 0.10$). A deeper analysis of metallicity dependence
on pulsation properties of Cepheids is in progress.


\begin{acknowledgement}
Financial support from the OTKA grant T046207 is gratefully
acknowledged.
\end{acknowledgement}



\begin{thebibliography}{}

\bibitem[Andrievsky et~al.(2002a)]{Andr-etal:A&A02a}
Andrievsky, S.~M., Kovtyukh, V.~V., Luck, R.~E., et~al. 2002a, A\&A, 381, 32

\bibitem[Andrievsky et~al.(2002b)]{Andr-etal:A&A02b}
Andrievsky, S.~M., Bersier, D., Kovtyukh, V.~V., et~al. 2002b, A\&A, 384, 140

\bibitem[Andrievsky et~al.(2002c)]{Andr-etal:A&A02c}
Andrievsky, S.~M., Kovtyukh, V.~V., Luck, R.~E., et~al. 2002c, A\&A, 392, 491

\bibitem[Andrievsky et~al.(2004)]{Andr-etal:A&A04}
Andrievsky, S.~M., Luck, R.~E., Martin, P., et~al. 2004, A\&A, 413, 159

\bibitem[Andrievsky et~al.(2005)]{Andr-etal:AJ05}
Andrievsky, S.~M., Luck, R.~E., and Kovtyukh, V.~V. 2005, \aj, 130, 1880

\bibitem[Berdnikov et~al.(2000)]{Berd-etal:a&as00}
Berdnikov, L.~N., Dambis, A.~K., and Vozyakova, O.~V. 2000, A\&AS 143, 211

\bibitem[Eggen(1985)]{Eggen:AJ85}
Eggen, O.~J. 1985, \aj, 90, 1278

\bibitem[Eichendorf and Reinhardt(1977)]{Eichen-etal:A&A77}
Eichendorf, M., and Reinhardt, M. 1977, A\&A, 61, 827

\bibitem[Fernie(1979)]{Fernie:PASP79}
Fernie, J.~D. 1979, \pasp, 91, 67

\bibitem[Fernie et~al.(1995)]{Fernie-etal:IBVS95}
Fernie, J.~D., Beattie, B., Evans, N.~R., and Seager, S. 1995, IBVS 4148

\bibitem[Fry and Carney(1997)]{Fry-etal.:AJ97}
Fry, A.~M., and Carney, B.~W. 1997 \aj, 113, 1073

\bibitem[Giridhar(1986)]{Giri:JApA86}
Giridhar, S. 1986, JApA, 7, 83

\bibitem[Groenewegen et~al.(2004)]{Groene-etal:A&A04}
Groenewegen, M.~A.~T., Romaniello, M., Primas, F., et~al. 2004, A\&A, 420, 655

\bibitem[Jurcsik and Kov\'acs(1996)]{Jurcsik-etal:A&A96}
Jurcsik, J., and Kov\'acs, G. 1996, A\&A, 312, 111

\bibitem[Kov\'acs and Zsoldos(1995)]{Kovacs-etal:A&A95}
Kov\'acs, G., and Zsoldos, E. 1995, A\&A, 293, 55L

\bibitem[Kovtyukh et~al.(2005)]{Kovty-etal:AJ05}
Kovtyukh, V.~V., Andrievsky, S.~M., Belik, S.~I., et~al. 2005, \aj, 129, 433

\bibitem[Kraft(1960)]{Kraft:ApJ60}
Kraft, R.~P. 1960, \apj, 132, 404

\bibitem[Luck et~al.(2003)]{Luck-etal:A&A03}
Luck, R.~E., Gieren, W.~P., Andrievsky, S.~M., et~al. 2003, A\&A, 401, 939

\bibitem[Sza\-ba\-dos(1997)]{Szabados:97}
Szabados, L. 1997, DSc Thesis

\bibitem[Zsoldos(1995)]{Zsoldos:ASPC95}
Zsoldos, E. 1995, ASPC, 83, 351


\end{thebibliography}
\end{document}